\documentclass[pra,twocolumn,superscriptaddress]{revtex4-1}
\usepackage{amsmath}
\usepackage{amssymb}
\usepackage{xfrac} 
\usepackage{amsmath,amssymb}
\usepackage[usenames]{color}
\usepackage{amssymb}
\usepackage{grffile}\include{learn_process_tensor_DPKM_GC_v1.tex}
\usepackage{amsmath, amstext, amssymb, amsfonts, amsxtra}
\usepackage{textcomp}
\usepackage{xspace} 
\usepackage[colorlinks]{hyperref}
\usepackage{tikz}
\usepackage{nicefrac} 
\usepackage[normalem]{ulem} 

\newcommand{\Hop}{H} 
\newcommand{\Dop}{\mathcal{D}}
\newcommand{\Lop}{\mathcal{L}}

\newcommand{\sgx}{\sigma^x}
\newcommand{\sgy}{\sigma^y}
\newcommand{\sgz}{\sigma^z}

\newcommand{\rhoop}{\rho}
\newcommand{\Xop}{X}
\newcommand{\Yop}{Y}
\newcommand{\Yest}{\hat{Y}}
\newcommand{\Yphys}{\mathcal{Y}}

\newcommand{\ExMPOup}{\Upsilon}

\newcommand{\MPOup}{\hat{\Upsilon}}
\newcommand{\tr}{{\rm tr}}
\newcommand{\evo}{\mathcal{E}}
\newcommand{\im}{{\rm i}}
\newcommand{\fidelity}{\mathcal{I}}
\newcommand{\ftnt}[1]{{\textsuperscript{\scriptsize\footnote{#1}}}}
\newcommand{\asf}{\mathsf{a}}

\newcommand{\timestep}{\delta t}

\begin{document}

\title{Tensor network based machine learning of non-Markovian quantum processes}

\author{Chu Guo} 
\affiliation{Henan Key Laboratory of Quantum Information and Cryptography, Zhengzhou, Henan 450000, China} 

\author{Kavan Modi}
\email{kavan.modi@monash.edu}
\affiliation{School of Physics and Astronomy, Monash University, Victoria 3800, Australia}
\affiliation{Institute for Quantum Science and Engineering, and Department of Physics, Southern University of Science and Technology, Shenzhen 518055, China}

\author{Dario Poletti} 
\affiliation{Science, Mathematics and Technology Cluster and Engineering Product Development Pillar, Singapore University of Technology and Design, 8 Somapah Road, 487372 Singapore}

\begin{abstract}
We show how to learn structures of generic, non-Markovian, quantum stochastic processes using a tensor network based machine learning algorithm. We do this by representing the process as a matrix product operator (MPO) and train it with a database of local input states at different times and the corresponding time-nonlocal output state. In particular, we analyze a qubit coupled to an environment and predict output state of the system at different time, as well as reconstruct the full system process. We show how the bond dimension of the MPO, a measure of non-Markovianity, depends on the properties of the system, of the environment and of their interaction. Hence, this study opens the way to a possible experimental investigation into the process tensor and its properties.
\end{abstract}

\date{\today}
\maketitle

A quantum machine, immersed in an environment, will undergo a quantum stochastic process. Such processes are usually highly complex and exhibit multi-scale temporal correlations. These correlations are often dubbed as \emph{memory} or \emph{non-Markovian} effects. Until recently, describing stochastic quantum processes and corresponding non-Markovian phenomena has been conceptually challenging and the traditional attempts had limited success~\cite{arXiv:1206.5346}. The fundamental reason for the disconnect between classical and quantum theories of stochastic processes is due to the invasive nature of quantum measurements. A classical stochastic process is the joint probability distribution of time-ordered random events, i.e., $p(\asf_{N}, \dots, \asf_0)$, where $\{\asf_{n}\}$ are the elements of the event space $\mathsf{A}_{n}$ at the $n$-th time step. Knowledge of such distribution allows to better predict the future given the observed events of the past. In contrast, a quantum event necessarily implies disturbance for the system. This poses a fundamental problem as a stochastic process in quantum physics needs to account for the effects of an unknown environment on a system which is, concurrently, affected by measurements.

To overcome these difficulties Refs.~\cite{PollockModi2018, arXiv:1512.07106} independently proposed the \emph{process tensor framework} to describe \emph{any} quantum stochastic process. Starting with the fact that a quantum event $\alpha$ corresponds to completely positive (CP) maps $\Lambda^\alpha$ belonging to an instrument $\mathcal{J} = \{\Lambda^\alpha\}$ that form a completely positive and trace preserving (CPTP) map $\Lambda = \sum_\alpha \Lambda^\alpha$. The probability of observing a sequence of events $\alpha_{0\rightarrow N} :=\{\alpha_{N}, \dots, \alpha_0 \}$, for a choice of a sequence of instruments $\mathcal{J}_{0\rightarrow N} := \{ \mathcal{J}_{N}, \dots, \mathcal{J}_0 \}$, is
\begin{gather}\label{eq:PT}
p(\alpha_{0\rightarrow N} |\mathcal{J}_{0\rightarrow N}) \!=\! \mbox{Tr}[\Lambda_N^{\alpha_N} \evo_{\timestep} \cdots \Lambda_1^{\alpha_1} \evo_{\timestep} (\rhoop_{0}^{\alpha_0} \!\otimes\! \rhoop^{E}_0)].
\end{gather}
Here, $\evo_{\timestep}$ are CPTP maps that deterministically evolve system-environment ($SE$) with action $\evo(\rhoop) = \sum_k E_k \rhoop E^\dag_k$. In this Letter, we assume that the initial $SE$ state is uncorrelated; We can use the techniques of Ref.~\cite{arXiv:1011.6138} to accommodate for initial $SE$ correlations.

\begin{figure}
\includegraphics[width=\columnwidth]{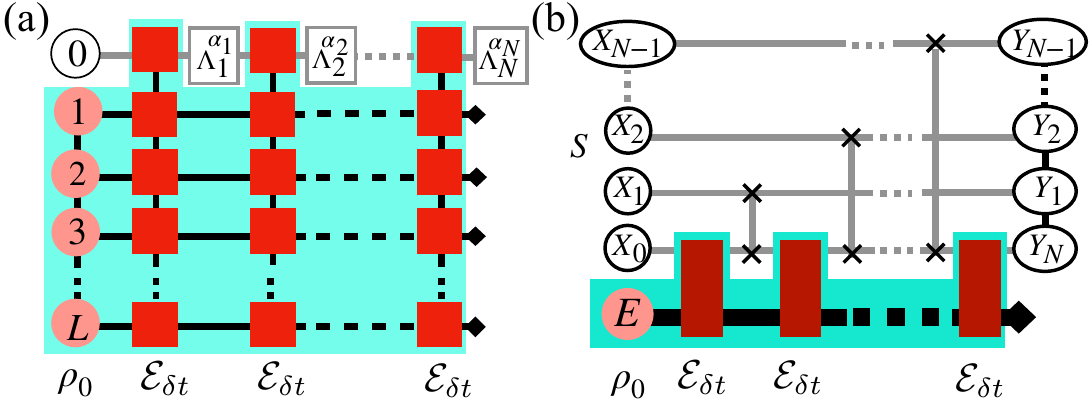}
\caption{(a) A many-body system-environment ($SE$), with the system ($S$) at site $0$ and the environment ($E$) of sites $1\rightarrow L$, the exact process tensor is obtained by contracting from bottom to top. Here, the initial $SE$ state $\rhoop_0$ is a matrix product state (MPS) and the $SE$ dynamics are due to the evolutionary operator $\evo_{\timestep}$, which is represented as a matrix product operator (MPO). The whole $SE$ process is represented as a two-dimensional tensor network where the horizontal dimension is the time direction and the vertical dimension is the spatial direction. The probability to observe any sequence of events $\Lambda_{0 \rightarrow N}$ for this process is given by Eq.~\eqref{eq:born}. (b) For an unknown $N-$step process, the process tensor can be experimentally reconstructed by replacing the $\Lambda$s by a set of input product states $X_{0\rightarrow N-1} = X_0 \otimes \dots \otimes X_{N-1}$ and corresponding output state $Y_{1\rightarrow N}$, which is a correlated state. This results in the green comb-like shape on the bottom, where the details of $E$ can be compressed at a desired cutoff.}\label{fig:evolution} 
\end{figure}

The last equation is reminiscent of a classical stochastic process, but the measurement and the process remain tangled. The untangling is achieved by a rearrangement
\begin{gather}\label{eq:born}
\left< \Lambda_{0\rightarrow N} \right>_{\Upsilon} = \mbox{Tr}[\Upsilon \ \left(\Lambda^{\alpha_{0\rightarrow N}}_{0\rightarrow N}\right)^{\text{T}}],
\end{gather}
where $\text{T}$ denotes transpose, $\Lambda^{\alpha_{0\rightarrow N}}_{0\rightarrow N} := \Lambda_N^{\alpha_N} \otimes \cdots \otimes \rho_0^{\alpha_0}$, and $\Upsilon := \mbox{Tr}_E [\evo_{\timestep} \star \cdots \star \evo_{\timestep} \rhoop^{E}_0]$, where $\star$ denotes the link product, defined in Ref.~\cite{arXiv:0904.4483}, which is a matrix product on the space $E$ and a tensor product on space $S$. Here, the process tensor $\Upsilon$ only depends on the environment and thus describes the process independent of the measurements. We depict the untangling in Fig.~\ref{fig:evolution}, where $\Lambda_{0\rightarrow N}$ and $\Upsilon_{0\rightarrow N}$ are seen as `comb'-like structures.

The process tensor is formally shown to be the quantum generalization of classical stochastic processes in Ref.~\cite{arXiv:1712.02589} and in the appropriate limit, it reduces to a classical stochastic process~\cite{PhysRevA.100.022120, arXiv:1907.05807}. As such, the process tensor is a powerful tool to study the dynamics of an open quantum system, especially when the environment has a finite memory size~\cite{PollockModi2018b, PhysRevLett.122.140401, PhysRevA.99.042108}. Importantly, a faithful description of a quantum process allows us to better predict the future behaviour of a system, given information about the past~\cite{arXiv:1902.07980}. However, in general, reconstructing the whole process tensor is hard. This is because its complexity grows exponentially in the number of time-steps.\ftnt{An $N$-step process tensor is a multipartite density operator on $2N-$partite Hilbert spaces. This is the natural the generalization of a $N$-partite probability distribution. The factor of 2 stems from the causal structure of quantum processes~\cite{arXiv:1712.02589}.}

In this Letter, we show that one can use the tensor-networks-based machine learning algorithm~\cite{GuoPoletti2018} to learn the outputs of the process tensor, as well as the full process tensor. To do this we exploit the fact a process tensor has a natural representation as an MPO~\cite{PollockModi2018} with the bond being a clear indication of the non-Markovian memory, i.e., the thick line inside the green region in Fig.~\ref{fig:evolution}(b). We train the process-MPO with data from the preparation of the states of $S$ at different times and corresponding measurement of local density operator of $S$. In particular, we analyze a qubit coupled to an environment and we follow its evolution as we prepare and measure it at different times. Our results open the way to experimentally reconstruct the process tensor. 

\textbf{Computing an exact process-MPO} --- We begin with an $L+1-$site $SE$ state $\rho_0$ in its MPS representation. The initial $SE$ state is evolved in discrete time steps $\timestep$ by successively applying the evolutionary operator $\evo_{\timestep}$ resulting in the two-dimensional tensor network shown in Fig.~\ref{fig:evolution}(a). To compute efficiently the process tensor from $\evo_{\timestep}$, we eliminate the environment degrees of freedom by contracting the two-dimensional tensor network from the bottom row to the top leading to an MPO with only the degrees of freedoms of $S$ left open.

From Fig.~\ref{fig:evolution}, we can readily estimate the complexity of the process tensor in terms of computer memory. Given the size of the inputs $d_X$ and that of the outputs $d_Y$,\ftnt{Here, for a single spin-$\sfrac{1}{2}$, $d_X=d_Y=4$.} the size of process tensor scales exponentially in the number of time steps as $(d_X d_Y)^{N}$. However, memory in realistic physical non-Markovian processes does not grow exponentially and should be bounded by the part of $E$ that stores information of system's past. In other words, while the process tensor grows with the number of time steps, the non-Markovian memory does not. Therefore, we can compress the process tensor to an MPO $\MPOup = \sum_{a_0,\dots a_{N}} W_{a_0, a_1}^{x_0, y_1} \cdots W_{a_{N-1}, a_{N}}^{x_{N-1}, y_N}$, where $x_{n}$ and $y_{n}$ are the physical indices, and $a_{n}$ are the indices for the ancillary system whose number is bounded by the MPO bond dimension $D$. The result is that the process tensor can be stored with computer memory that goes as $N d_X d_Y D^2$. Clearly, for Markovian dynamics the bond dimension needed to accurately represent the process tensor would be $D=1$.

\textbf{Learning an approximate process-MPO} --- In practice, one may not have access to the evolutionary operator $\evo_{\timestep}$. Instead, one can prepare the quantum state and perform measurements on the quantum state, i.e., quantum process tomography~\cite{Milz_OSID}. However, as argued above, this procedure grows exponentially both with the number of spins in $S$ and the number of time steps $N$. A way around this is to accurately learn the process tensor with machine learning techniques based only on the data of a limited number of input-output pairs. In particular, in~\cite{GuoPoletti2018} they showed that given a list of input-output MPS pairs, it is possible to efficiently learn an optimal MPO as a mapping between those pairs, by using a variational MPO ansatz and an iterative procedure which closely resembles the variational matrix product states method used to solve for the ground state of one-dimensional quantum states. This method has proven to be able to successfully predict the evolution of different dynamical systems, from cellular automata to non-linear diffusion equations~\cite{GuoPoletti2018}, even performing better than bidirectional \textit{long short-term memory} recurrent neural networks~\cite{lstm}, a state of the art model in natural language processing.

Let $\Xop_{0\rightarrow N-1}^{m}$ and $\Yop_{1\rightarrow N}^{m}$ be the inputs and outputs of the process tensor, as depicted in Fig.~\ref{fig:evolution}(b). These states can be thought of as a list of MPS pairs corresponding to the density operators $\Xop_{0\rightarrow N-1}^{m}$ and $\Yop_{1\rightarrow N}^{m}$. Here, $m$ is the labelling for the training (testing) data that runs from $1$ to $M_{train}$ ($M_{test}$) for training (testing) phases. Let the mapping from $\MPOup:\vert \Xop_{0\rightarrow N-1}^{m} \rangle \longrightarrow \vert \Yest_{1\rightarrow N}^{m} \rangle$ be parameterized by our MPO ansatz $\MPOup$, where we use the symbols with hat $\MPOup$ and $\Yest^m_{1\rightarrow N}$ for quantities which we learn. We define the mean square loss function as $f(\MPOup) = \sum_{m=1}^M \left< \Delta \Yop_{1\rightarrow N}^{m}  | \Delta \Yop_{1\rightarrow N}^{m} \right> + \mu \, \tr(\MPOup^{\dagger}\MPOup)$, where $\left| \Delta \Yop_{1\rightarrow N}^{m} \right> := \vert \Yop_{1\rightarrow N}^{m} \rangle - \vert \Yest_{1\rightarrow N}^{m} \rangle$ and the term $\mu$ is for regularization. As done in~\cite{GuoPoletti2018}, and similar to variational matrix product states method~\cite{schollwock2011}, the minimum of the loss function above can be obtained by iteratively solving for the local minimum via $\partial f(\MPOup)/ \partial W_{a_{n-1}, a_{n}}^{x_{n-1}, y_{n}} = 0$. We use ten sweeps for the optimization of the loss function for all the simulations, where the loss function decreases most significantly in the first two sweeps.

\textbf{Model} --- To demonstrate the performance of our method, we apply it to a dissipative quantum spin chain of $L+1$ spins as shown in Fig.~\ref{fig:evolution}(b). Here, site 0 is the $S$ and the rest are the $E$, which can be strongly coupled to the system. The total $SE$ Hamiltonian $\Hop = H_S + H_E + H_{int}$, where
\begin{align}\label{eq:tot_Ham} 
&H_S = h \sigma^z_0, \quad H_{int} = J  \sum_{l=1}^{L} \left(\sgx_{0}\sgx_l + \sgy_{0}\sgy_l\right)\\
&H_E = h \sum_{l=1}^{L}\sgz_l + J_E \sum_{l=1}^{L-1} \left(\sgx_{l} \sgx_{l+1} + \sgy_{l} \sgy_{l+1} + \Delta\sgz_{l} \sgz_{l+1} \right). \nonumber
\end{align}
Above, $J$ is the coupling between the system and each of the environment spins, $h$ is the magnetization strength, $\Delta$ is the anisotropy for the environment. We work in the units where both $J_E=1$ and $\hbar=1$. 

To consider a more generic scenario, the environment spins are also subjected to a uniform dissipation which can be described by a dissipator in Gorini-Kossakowski-Sudarshan-Lindblad form~\cite{GoriniSudarshan1976, Lindblad1976}
\begin{gather}\label{eq:dissipator}
\Dop(\rhoop) = \gamma \sum_{l=1}^{L} \sum_{j=\pm} \nu_j \left(2\sigma^j_l \rhoop \ \sigma^{j\dag}_l - \{ \sigma^{j\dag}_l\sigma^j_l, \rhoop\} \right),
\end{gather}
where $\gamma$ is the dissipation strength,
$\nu_+ = r$ and $\nu_- = (1 - r)$ with $r$ being the average occupation. The $SE$ dynamics is thus described by the master equation, $\dot{\rhoop} = \Lop(\rhoop) =  -\im \left[\Hop, \rhoop\right] + \Dop(\rhoop)$. Such a scenario, in which a quantum system coupled to an environment is modeled by the system being coupled to an extended system (in this case $L$ spins) which are then coupled to a bath, has been increasingly used in the field of quantum thermodynamics~\cite{GargAmbegaokar1985, GelbwaserAspuruGuzik2015, IlesSmithNazir2014, NewmanNazir2017, StrasbergEsposito2018, Nazir}. In the following we choose the state of $E$ to be $\rhoop^E_0 = \left(r\vert 0\rangle\langle 0\vert + (1-r)\vert 1\rangle\langle 1\vert\right)^{\otimes L}$, which is the steady state of the dissipator Eq.~\eqref{eq:dissipator}.

We generate a product input $X_{0 \to N-1}^{m}$ and the corresponding correlated output $Y_{1 \to N}^{m}$ to train (and to test) $\hat{\Upsilon}$. To prepare the training and test data we evolve the $SE$ with the master equation at discrete time steps $n\delta t$, where $n=0,\dots,N$ and the output density matrix $\Yop_{1 \to N}^{m}$ is evaluated. 

For testing, we only consider the fidelity of local output states $\Yop_{n}^{m}$. However, in the testing phase, we find that the predicted local density operator $\Yest^{m}_{n}$ is not always positive definite, therefore we replace it by the corresponding closest, with respect to norm-2, physical density matrix $\Yphys^{m}_{n}$ in the Bloch sphere. By means of quantum fidelity, $\mathcal{F} (a,b) := [\tr(\sqrt{\sqrt{a} \, b \sqrt{a}})]^2$, we compare $\Yop^{m}_{n}$ to $\Yphys^{m}_{n}$ for each testing set element $m$ and time step $n$ to evaluate the quality of our trained process tensor. Below, the learning performance is reported by means of the median of the infidelity over $M_{test}$ and $N$, i.e., $\fidelity = {\rm med}_{n} \left(\fidelity_{n}\right)$ with $\fidelity_{n} = {\rm med}_{M_{test}} \left(1-\mathcal{F} (\Yphys^{m}_{n}, \Yop^{m}_{n})\right)$.

\begin{figure}
\includegraphics[width=.9\columnwidth]{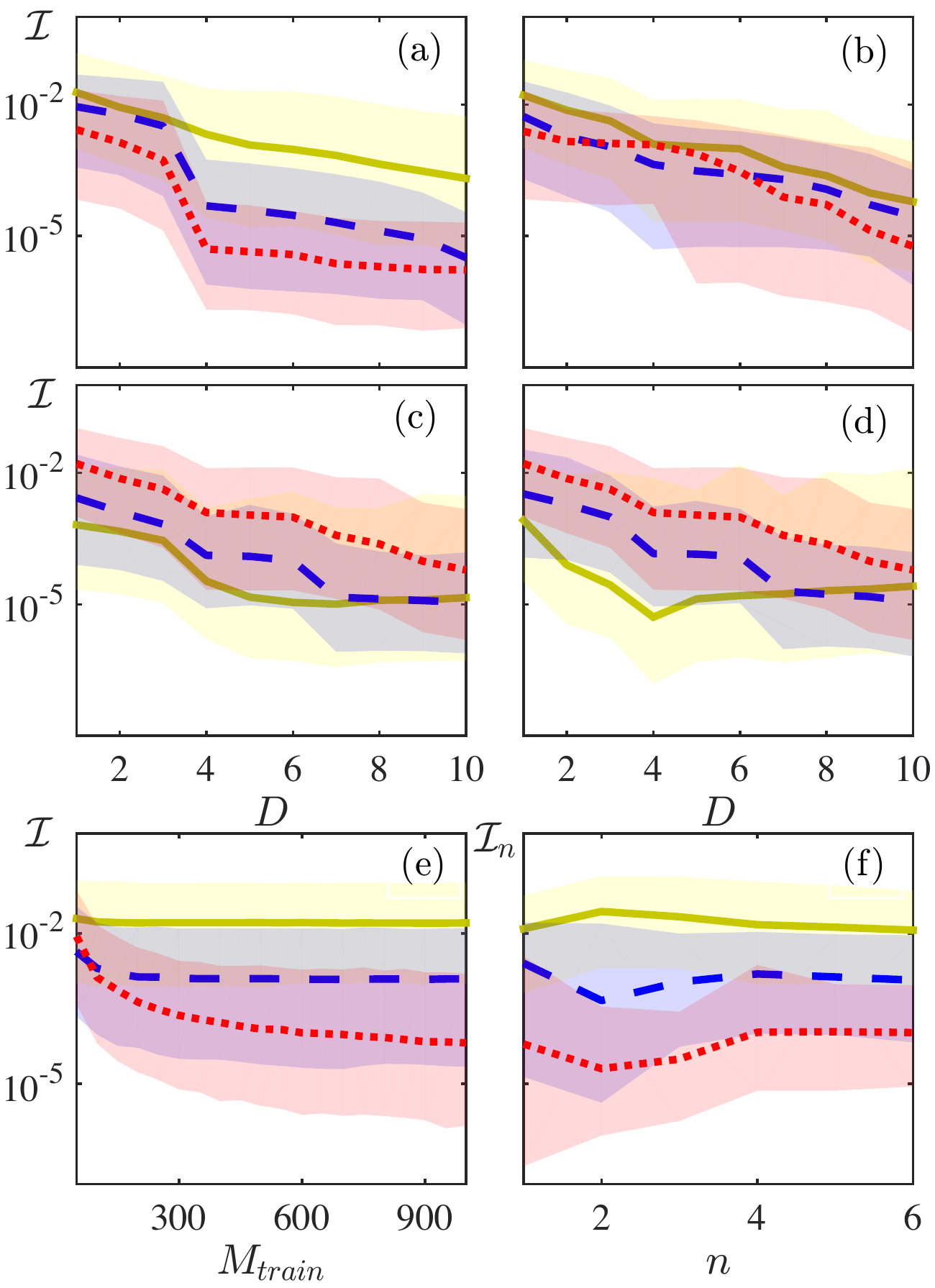}
\caption{(a-d) Infidelity $\fidelity$ as a function of bond dimension $D$. The solid, dashed, dotted lines correspond to (a) $\gamma=0, 5, 10$, and (b) $r=0,0.25,0.5$, (c) $J=1,2,4$ and (d) $\delta t=0.01, 0.05, 0.1$ respectively. (e) Infidelity $\fidelity$ as a function of number of training samples $M_{train}$, and (f) time-resolved infidelity $\fidelity_{n}$ as a function of the time step $n$. In (e,f) the solid, dashed, dotted lines correspond to $D=1,4,10$ respectively. The default parameters used in all the panels, unless otherwise specified, are $N=6$, $\delta t=0.1$, $M_{train}=1000$, $M_{test}=500$, $J=4$, $\Delta=1.5$, $h=0.5$, $\gamma=1$, $\Delta=1.5$, $h=0.5$, $r=0$. In all panels the shaded area corresponds to the $95\%$ confidence level of the curve in the same color. }\label{fig:vs_J} 
\end{figure}

\textbf{Numerical results} ---
For our numerical simulations, we consider $L=3$ and $N=6$, which amounts to a simulation of $9$ spins, $L+1$ for the initial $SE$ state and $N-1$ for each remaining time step, see Fig.~\ref{fig:evolution}(b). We first study the quality of predicted density matrices $\Yphys^{m}_{n}$ against different parameters in the master equation. We first consider the parameters of $E$, i.e., the dissipation rate $\gamma$ and the average occupation $r$. In panel (a), we observe the average infidelity versus bond dimension $D$ for different dissipation rates $\gamma$. For larger $\gamma$, e.g. $\gamma=10$, the state of $E$ resets quickly and thus a smaller bond dimension $D$ is sufficient to obtain small infidelities. In all panels of Fig.~\ref{fig:vs_J} the shaded areas correspond to the $95\%$ confidence interval of the curve with the respective color. Similarly, in Fig.~\ref{fig:vs_J}(b), we consider a dissipation rate $\gamma=1$, which requires larger bond dimensions, and we vary the average occupations $r$. As $r \to 0.5$, the memory required to obtain small infidelities is significantly reduced because the state of $E$ is closer to the infinite temperature state. In particular, a bond dimension $D=7$ is sufficient to obtain $\fidelity \approx 7.7 \times 10^{-5}$ for $r=0.5$. It is instead more difficult to learn the process tensor when $r=0$, namely when $E$ is set to a pure state of spins pointing down. In this case a bond dimension $D=10$ is needed to obtain low infidelities $\fidelity \approx 6\times 10^{-5}$.\ftnt{We note that the accuracy of $10^{-5}$ is also due to the time evolution method that we have implemented for the open dynamics from Eqs.~(\ref{eq:tot_Ham},\ref{eq:dissipator}).}

Two other important parameters to be considered are the $SE$ coupling strength $J$ and the time between observations $\delta t$. For the time scale considered, with $\delta t=0.1$ and $N=6$, larger interactions $J$ lead to stronger $SE$ coupling, thus requiring a larger bond dimension, in this case $D=9$ for $J=4$, to reach small infidelities $\fidelity\approx 9.6\times 10^{-5}$. In comparison, in case $J=1$, with a bond dimension $D=4$ one could already reach $\fidelity\approx 3.4\times 10^{-5}$. We then consider the case with large coupling $J=4$ and we analyze the effect of different observation time steps $\delta t$ in Fig.~\ref{fig:vs_J}(d). Here we observe that larger time steps allow to reach smaller infidelities compared to smaller ones. This is due to the fact that for larger times $\delta t$, the dissipator from Eq.~\eqref{eq:dissipator}, which here is taken with $\gamma=1$ and $r=0$, can more effectively remove memory from $E$. In both panels of Fig.~\ref{fig:vs_J}(c,d), we observe that for bond dimensions larger than $D\approx 7$, the fidelity does not improve significantly. We deem that this is an effect of the number of parameters that need to be trained, versus the number of training test. To study this in more detail, in Fig.~\ref{fig:vs_J}(e) we plot how the infidelity $\fidelity$ varies with the number of training sets $M_{train}$ for different bond dimensions. We observe that for smaller bond dimensions a smaller number of training samples is needed because one needs to train less parameters. However the performance are limited. To reach smaller infidelities one needs larger bond dimensions, e.g. $D=10$, and then a larger number of training data is required. Until now we have discussed the average infidelity, which takes into the overall infidelity $\fidelity$ for the different $N$ time steps. We thus also explore how the infidelity varies at each time step $n$. Because of the way we train the process tensor, which gives the same importance to each point in time, we do not see in Fig.~\ref{fig:vs_J}(f) a clear change of the infidelity versus the time steps $n$, and all the oscillations are within the error bars.

\begin{figure}
\includegraphics[width=0.9\columnwidth]{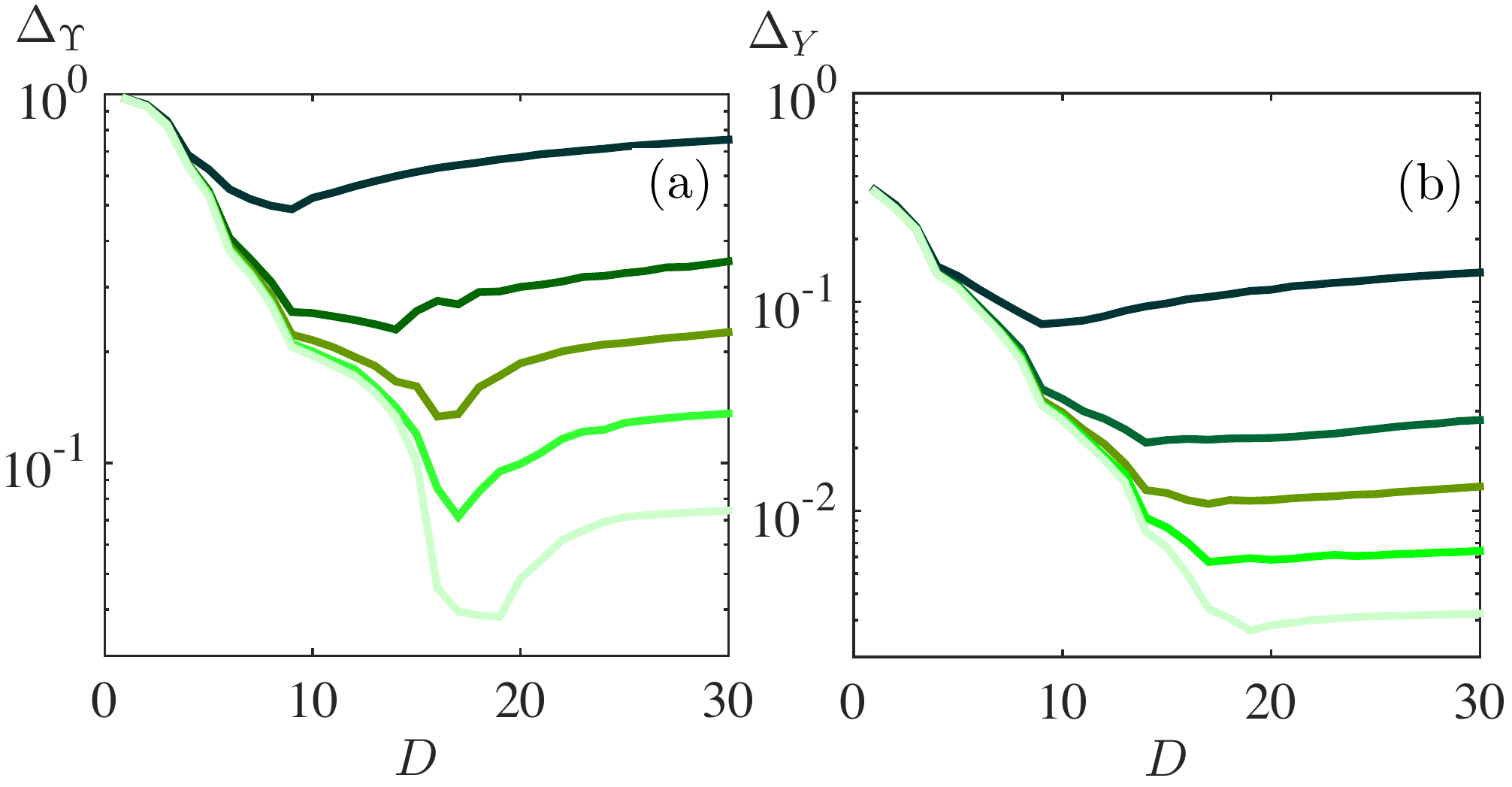}
\caption{(a) Distance $\Delta_{\Upsilon}$ between the exact process tensor and the trained process tensor with different bond dimensions. (b) Distance $\Delta_Y$ between the exact output and the predicted output averaged over $500$ testing data. In both figures the darker to lighter lines correspond to $M_{train}= 100, 500, 1000, 2000, 4000$ respectively. Other parameters used are $N=5$, $\delta t=0.1$, $J=4$, $\Delta=1.5$, $h=0.5$, $\gamma=1$, $\Delta=1.5$, $h=0.5$, $r=0$.}
\label{fig:fig3} 
\end{figure}

Now, instead of local observables, we directly compare the trained process $\MPOup$ with the exact one $\ExMPOup$. In Fig.~\ref{fig:fig3}(a) we plot, as a function of the bond dimension $D$ and for different number of training data $M_{train}$, the distance between $\ExMPOup$ and $\MPOup$, defined as $\Delta_{\Upsilon} = \| \ExMPOup \|^{-1} \|\MPOup(D) - \ExMPOup \|$, where the norm of a process tensor is computed by first reshaping the process tensor into a vector and then computing the vector $2-$norm. We can see that given enough number of training data and a large enough bond dimension $D$, the trained process tensor will converge to the exact one. The rising of the tails in Fig.~\ref{fig:fig3}(a) are due to the fact that the number of parameters of the MPO increases but the number of training data remains fixed. In Fig.~\ref{fig:fig3}(b), we evaluate the quality of the trained process-MPO $\MPOup$ by computing the distance between the overall predicted output $\vert\Yest^{m}_{1 \rightarrow N}\rangle$ and the exact output $\vert \Yop^{m}_{1 \rightarrow N}\rangle$. Unlike the local in time $\vert\Yest^{m}_{n}\rangle$, the operators $\vert\Yest^{m}_{1 \rightarrow N}\rangle$ and $\vert \Yop^{m}_{1 \rightarrow N}\rangle$ take into account correlations between different times. We thus define the distance between the two operators as $\Delta_Y = M_{test}^{-1} \sum_{l=1}^{M_{test}}\langle \Delta \Yop^{m}_{1 \rightarrow N} \vert \Delta \Yop^{m}_{1 \rightarrow N} \rangle$, where $\Delta Y^{m}_{1 \rightarrow N}$ is defined above Eq.~\eqref{eq:tot_Ham}. In Fig.~\ref{fig:fig3}(b) we observe, even more clearly than before, an exponential decrease of the distance with increasing bond dimension, provided a large enough number of training data is used, This is due to the increased representability of the MPO with increasing bond dimension D. The increase of distances $\Delta_{\Upsilon}$ and $\Delta_Y$ are due to keeping $M_{test}$ constant while the number of parameters of the MPO increases. Finally, our method complements Refs.~\cite{arXiv:1901.05158, arXiv:1902.07019}, which use machine learning to estimate the size of the bond of the process tensor and non-Markovian dynamics, respectively. However, our method directly estimates the process tensor itself.

\textbf{Discussions} --- 
In this work, we have used a tensor network-based machine learning algorithm to learn a non-Markovian quantum process. Specifically, we trained an MPO to predict the output states of a multi-time quantum process, the full characterization of the open dynamics of a quantum system, and also local in time properties. We note that, unlike standard process tomography procedure~\cite{PollockModi2018, Milz_OSID}, which requires preparing a specific set of basis states for the input, we rely on random input states. While here we used the output density matrices $Y^m_{1 \to N}$ to learn the process, we can just as well learn the process from a set of random local preparations followed by a set of local measurements. The power of our learning algorithm is most notable when used to learn partial information of the process tensor, for example, predicting measurement outcomes in a specific measurement basis. Adding efficient tomography technique, such as the MPS tomography~\cite{Cramer2010, MPStomexp}, will speed up the process reconstruction in comparison to quantum process tomography. An interesting future avenue will be to combine our learning algorithm with existing tensor network methods, designed to solve specific non-Markovian dynamical problems, such as the e.g. the spin-boson model~\cite{tempo, arXiv:1812.00043, arXiv:1902.00315}, or to use it to uncover patterns in non-Markovian memory~\cite{arXiv:1704.00800, Pollock2018T, arXiv:1811.03722, arXiv:1907.12583}. Our method can also allow to reconstruct processes for experiments that have limited control~\cite{PhysRevA.98.012108, ibmrestricted} (e.g. in today's quantum computers intermediate measurements are not possible). Finally, our tensor network based approach is compatible with quantum machine learning algorithms and can directly be adapted to work with input and output quantum states.

\begin{acknowledgments}
This research was initiated at the Kavli Institute of Theoretical Physics in Santa Barbara, California and was supported in part by the National Science Foundation under Grant No. NSF PHY-1748958. CG acknowledges support from National Natural Science Foundation of China under Grants No. 11805279. DP acknowledges support from Ministry of Education of Singapore AcRF MOE Tier-II (project MOE2018-T2-2-142). KM is supported through Australian Research Council Future Fellowship FT160100073.  The computational work for this article was partially performed on resources of the National Supercomputing Centre, Singapore (NSCC)~\cite{nscc}. 
\end{acknowledgments}

\bibliography{refs}
\end{document}